\documentclass[twocolumn,showpacs,preprintnumbers,amsmath,amssymb]{revtex4}
\usepackage{graphicx}
\usepackage{amssymb}
\usepackage[latin1]{inputenc}
\usepackage{dcolumn}
\usepackage{bm}

\begin{document}


\title{Optical  properties of  magnetically doped ultra-thin topological insulator slabs. }

\author{Martha Lasia and Luis Brey}
\affiliation{Instituto de Ciencia de Materiales de Madrid, (CSIC),
Cantoblanco, 28049 Madrid, Spain }

\date{\today}

\keywords{Topological insulators \sep Electronic properties \sep
Transport properties}
\pacs{72.25.Dc,73.20.-r,73.50.-h}

\begin{abstract}
Starting from a three dimensional Hamiltonian, we study the optical properties of ultra-thin topological insulator  slabs  for which the coupling between Dirac fermions on opposite surfaces results
in two degenerated gapped hyperbolic bands.  
The gap is a threshold  for the optical absorption and translates in a peak in the imaginary part of the optical
conductivity.
An exchange field applied perpendicular to the slab splits the degenerated hyperbolic bands and a double step structure
come out  in the optical absorption, whereas  a double peak structure appears 
in the
imaginary part of the longitudinal optical conductivity.  The exchange field breaks time-reversal symmetry and 
for exchange fields larger than the surfaces coupling  gap, the zero frequency Hall conductivity is quantized to $e^2/h$. This result implies  large values of the
Kerr and Faraday rotation angles.  
In ultra-thin slabs, the absence of  light multiple scattering and bulk conductivity,   makes  the Kerr and Faradays angles  to remain  rather  large in a  wide range of frequencies.

\end{abstract}
\maketitle

\section{Introduction.}
Three dimensional (3D) topological insulators (TI's) are materials that possess  conducting metallic surface states in the bulk energy gap\cite{hasan_2010,Qi_2011,YAndo_2013}. These systems typically are band insulators where  strong spin orbit coupling alters  the natural 
order in energies  of the band structure.  At the surface of the TI, in contact with the vacuum,  the bands turn back to their natural  order, and a two-dimensional
metallic state merges at the surface. These surface states have an helical linear dispersion and the quasiparticles are governed by a two-dimensional (2D) massless Dirac equation. The Dirac cone is centered at a time reversal invariant point in the two dimensional Brillouin zone, and the degeneracy at the Dirac point  is preserved by time reversal symmetry.  

At zero temperature the optical conductivity of a system described by the two-dimensional  Dirac Hamiltonian, with the chemical potential located at the Dirac point, gets a universal value of $\sigma _0$=$\frac {\pi} 2 \frac {e^2} h$.  Light transmittance experiments  have confirmed this universal value for the optical absorption in graphene\cite{Nair_2008,Tobias_2008}.  
The optical properties of TI surface states has been theoretically studied recently\cite{Li_2013,Xiao_2013,Schmeltzer_2013,Peres_2013},
and the  same universal value for  the optical conductivity, $\sigma _0$, has been obtained\cite{Schmeltzer_2013,Peres_2013}.
However,   hexagonal warping terms to appear in the surface  band structure of TI's  modify the interband optical transition and deviations from the universal background as seen in graphene\cite{Nair_2008,Tobias_2008} have been predicted\cite{Li_2013,Xiao_2013}. 
The optical conductivity of Bismuth based topological insulators has been experimentally studied\cite{LaForge_2010,DiPietro_2012,Akrap_2012,Post_2013}, and the main result is that the exotic properties of the TI surface states are masked by high carrier densities and impurity band conduction.

The magnetoelectric coupling that occurs at the surfaces  of topological insulators\cite{hasan_2010,Qi_2011,YAndo_2013} is the origin of new and exotic  phenomena such as the possibility of inducing magnetic monopoles\cite{Qi_2009}, the tunable  Casimir effect\cite{Grushin_2011} and the giant magneto-optical Kerr and Faraday effects\cite{Qi_2008,Essin_2009,Tse_2010,Tse_2011,Karch_2009,Chang_2009}.
The Kerr and Faraday effects describe the rotation of the light polarization when  it is reflected or transmitted respectively by a magnetic material.
When time-reversal symmetry is broken at the surface of a TI, a gap is induced in the Dirac band structure and the surface shows  an anomalous Hall effect, $\sigma _{xy}$=$\frac 1 2 e^2 /\hbar$. In this situation  TI surfaces\cite{Qi_2008,Essin_2009} or a clean thin  slabs\cite{Tse_2010,Tse_2011}  will have  a strong magnetoelectric effect that manifests, in ideal systems, in a universal Faraday rotation and a  giant Kerr rotation.
At the surfaces  of TI's  the time reversal symmetry can be broken without applying external magnetic fields.
Doping the system with magnetic impurities induces  an exchange field acting on the TI surface state\cite{Chang_2013}.  Experiments in Bi$_2$Se$_3$  thin films indicate the existence of colossal Kerr\cite{Valdes_2012} and Faraday\cite{Jenkins_2012} rotations in the THz regime when time reversal symmetry is broken by a strong  magnetic field perpendicular to the slab.

Topological insulators as Bi$_2$Te$_3$, Bi$_2$Se$_3$ and (BiSb)Te, have a layered structure consisting of stacked quintuple layers with relatively weak coupling between them. Therefore, it is possible to prepare these crystals in the form of thin films. In thin films the bulk contribution to electrical conductivity and optical absorption can be reduced considerably, and these systems can be the appropriated geometry to observe surface properties of TI's\cite{Chang_2013,Wang_2013,Lu_2013}.  When the thickness of the thin film is of the order of the surface-state decay length into the bulk ($\sim 10$nm), the tunneling between the top and bottom surfaces opens an energy gap and two degenerate massive Dirac bands appear\cite{Linder_2009,Zhang_2010,Lu_2010,Liu_2010b}. 
By increasing the layer thickness the gap decreases and oscillates  and the system alternates between a trivial topological phase and a 2D quantum spin Hall topological phase. 
The existence of this gap indicates that the ultra-thin film behaves as a quasi 2D system, and not as a couple of  2D electron gases separated by a dielectric.

In this work we study the electrical and optical properties of ultra-thin TI slabs  in presence of an exchange field. Starting form a realistic 
4$\times$4 ${\bf k }\cdot {\bf p}$ Hamiltonian we compute the band structure and optical conductivity of TI slabs. 
The utilization of a realistic bulk Hamiltonian as the starting point releases   the use of an momentum or energy cutoff in the calculations. Also,  by  starting from the 3D Hamiltonian,  the surface states dispersion  contains automatically quadratic momentum  terms and electron hole asymmetry.  
We study the competition between the confining gap due to the coupling between states on opposite surfaces and the gap induced by the exchange field. 
Using the Kubo formula we calculate the optical conductivity of the TI slab.
The longitudinal conductivity gives  information of the optical absorption of the slab, whereas the zero frequency 
Hall conductivity indicates the topological  character of the system.  From the optical conductivity we obtain the Kerr and Faraday angles of a ultra thin TI slab. We obtain a giant Kerr angle and a quantized Faraday angle. Both angles get large values in a wide window of low frequencies and because the ultra-thin slab behaves as a 2D system they are not affected by TI  bulk conductivity or  by multiple reflection inside the TI slab.

\section{Bulk Hamiltonian}

The low energy band structure  of three dimensional topological
insulators in the Bi$_2$Se$_3$ family of materials can be
described by a four band Hamiltonian proposed by Zhang {\it et
al.} \cite{Zhang_2009,Liu_2010}. In the ${\bf k }\cdot {\bf p }$
formalism, states near zero energy and long wavelength
are governed by a Hamiltonian of the form
\begin{equation}
H^{3D}  =  E({\bf k}) +
\left(
  \begin{array}{cccc}
     \mathcal{M} ( {\bf k} ) & A_1 k_z & 0 & A_2 k _ - \\
    A _1 k_z   & -\mathcal{M}( {\bf k} ) & A_2 k _ - & 0 \\
    0 & A_2 k _ +  &   \mathcal{M} ( {\bf k} ) & - A _1 k _z \\
    A_2 k _ +  & 0 & -A _1 k _z  & - \mathcal{M}( {\bf k} ) \\
  \end{array}
\right),
\label{H3D}
\end{equation}
where $\mathcal{M} ( {\bf k} )$=$M_0 - B_2 (k_x ^2 + k_y ^2)-B_1k_z
^2$, $k_{\pm}$=$k_x \pm i k_y$ and  $E({\bf k})$=$ C + D _1 k_z ^2 +
D_2 (k_x ^2+k_y ^2)$. The four basis states for which this Hamiltonian is written are
$|1>$=$|p1 _z ^+, \uparrow>$, $|2>$=$-i |p2 _z ^- ,
\uparrow>$,$|3>$=$|p1 _z ^+, \downarrow>$, and $|4>$= $i|p2 _z ^-,
\downarrow>$, which are hybridized states of the Se $p$ orbitals
and the Bi $p$ orbitals,  with the superscripts  $(\pm )$ standing for even and odd parity, and 
$\uparrow$ and $\downarrow$ for spin up and down respectively. The Hamiltonian
parameters for a particular material can be obtained by fitting to density functional band structure calculations \cite{Liu_2010}. In the case of Bi$_2$Se$_3$
the relevant parameters are
$M_0$=0.28$eV$, $A_1$=2.2$eV \AA $, $A_2$=4.1$eV\AA$, $B_1$=10$eV \AA ^2$, $B_2$=56.6$eV\AA^2$,
$C$=-0.0068$eV$, $D_1$=1.3$eV \AA ^2$ and $D_2$=19.6$eV\AA^2$.

\subsection{Electronic Structure of  Topological Insulator Slabs.} 

We analyze  a TI slab perpendicular to the
$z$-direction and  thickness $L$ . The system is invariant in the $(x,y)$-plane so that $k_x$ and $k_y$ are good
quantum numbers.
The eigenvalues,  $\varepsilon _{n,{\bf k}}$, and
wavefunctions, $\Psi _{n, {\bf k}} (z)$, are obtained by solving
Eq.\ref{H3D} with $k_z = -i\partial _z$ and forcing the wavefunction
to vanish at the surfaces of the slab,  $z=0$ and $z=L$. This is satisfied expanding $\Psi _{n,
{\bf k}} (z)$ in harmonics,
of the form $\sin {(
\frac l L  \pi z)}$, being $l$ a positive integer\cite{Lasia_2013}. For a given
two-dimensional wavevector, $\bf k $ =$(k_x,k_y)$,  we diagonalize
the Hamiltonian in this basis and we obtain  a discrete number of
eigenvalues, $\varepsilon _{n,{\bf k}}$ and the corresponding
wavefunctions,
\begin{equation}
\Psi _{n, {\bf k}} (z)= \frac {e ^{i {\bf k } {\bf r}}} { \sqrt{A} }
\sqrt { \frac 2 L}\sum _{l =1} ^{N_{\mathrm{max}}}  \sum _{ j=1,4 }a
_{n,j} ^l ({\bf k}) \sin  {(l \pi  \frac z  L)} \,  |j>\, \, ,
 \label{2Dwf}
\end{equation}
here $A$ is the sample area. The number of harmonics used in the
expansion, $N_{max}$, depends on the thickness of the slab, and it
is chosen large enough so that the results do not depend on its
value. 

\begin{figure}
\includegraphics[clip,width=8.cm]{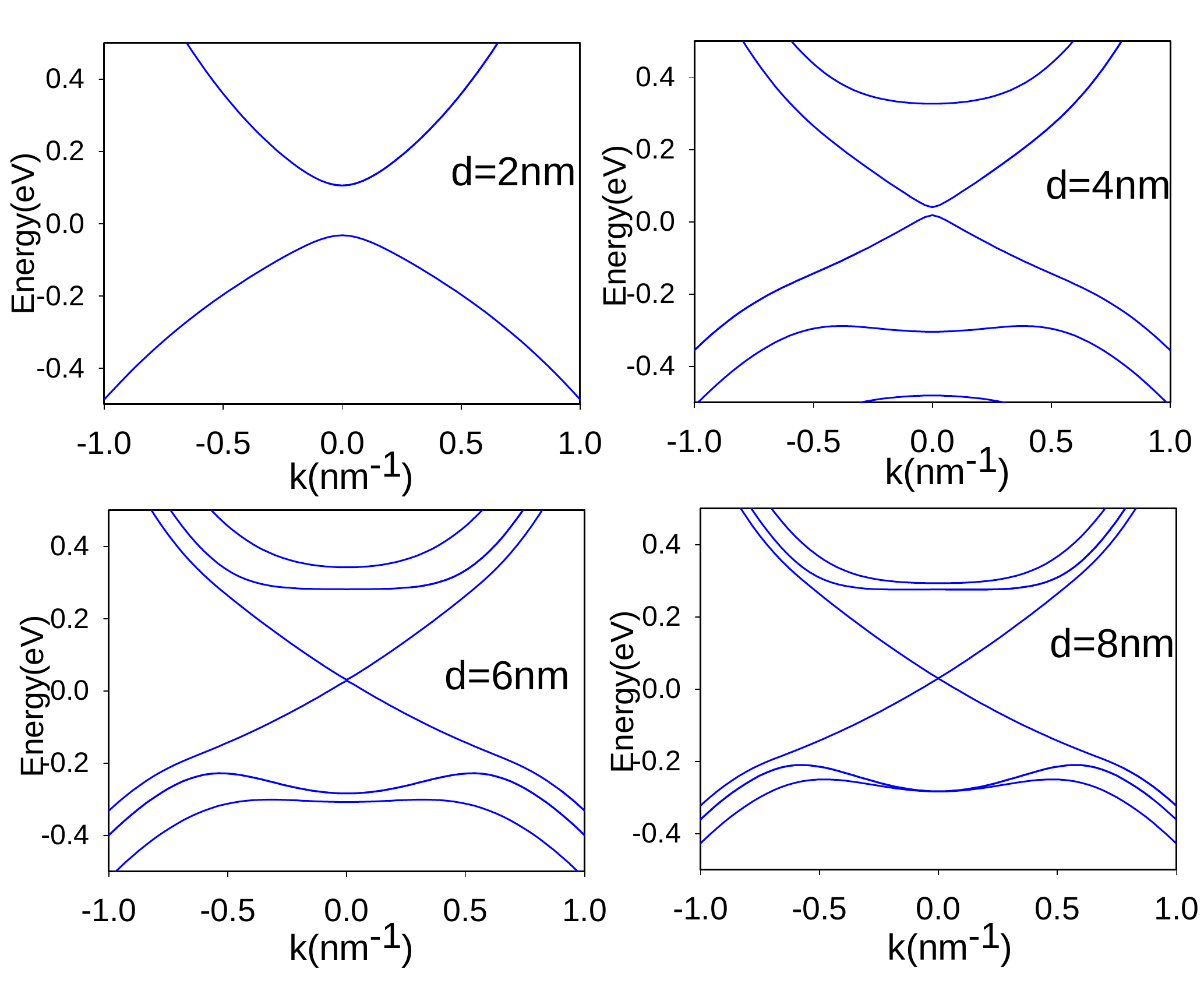}
\caption{(Color online) Band structure of TI slabs of different thickness $d$. The slabs are perpendicular    to the $\hat z$ direction. The electronic sates are  obtained by diagonalizing Hamiltonian  Eq. \ref{H3D} with the appropriated boundary conditions. At the Brillouin zone center, $\Gamma$,  the bulk gap has a value 2$M_0$=0.56eV. Energies inside the bulk energy gap correspond to surface states. For thick slabs low energy surface states dispersion has the form of a Dirac-cone. For thin slabs,   coupling between surface states located on opposite surfaces opens  an energy gap in the Dirac spectrum. The system has circular symmetry and we plot the bands as function of the absolute value of the in-plane wavevector ${\bf k}$.}
 \label{bandasB0}
\end{figure}

In Fig.\ref{bandasB0} we plot the band structure of thin TI slabs for different values of the  thickness $d$. 
For thick TI slabs ($d>6nm$) the surfaces are practically decoupled 
and  two degenerated gapless Dirac-like  bands, one for each surface, appear in the bulk energy gap region.   
This is the benchmark of the TI.
As the thickness of the slab decreases the electronic states localized on opposite surfaces couple 
and the Dirac cones transform in two degenerated  hyperbolic  Dirac bands with gap $E_g$ at the center of the Brillouin zone $\Gamma$.
In agreement with previous  works\cite{Linder_2009,Zhang_2010,Lu_2010,Liu_2010b} the values of this gap oscillates as  function of the thickness $d$. 
For each set of gapped Dirac bands,  the expectation value of the spin in the valence and conduction bands,  gets a ${\bf k}$-dependent configuration near the center of the Brillouin zone\cite{Liu_2010b,Lu_2010}. 
Because of  time reversal symmetry,  the other pair of 
Dirac hyperbolas is  a degenerated copy with the opposite  ${\bf k}$-dependent spin configuration\cite{Liu_2010b,Lu_2010}.

\begin{figure}
\includegraphics[clip,width=7.5cm]{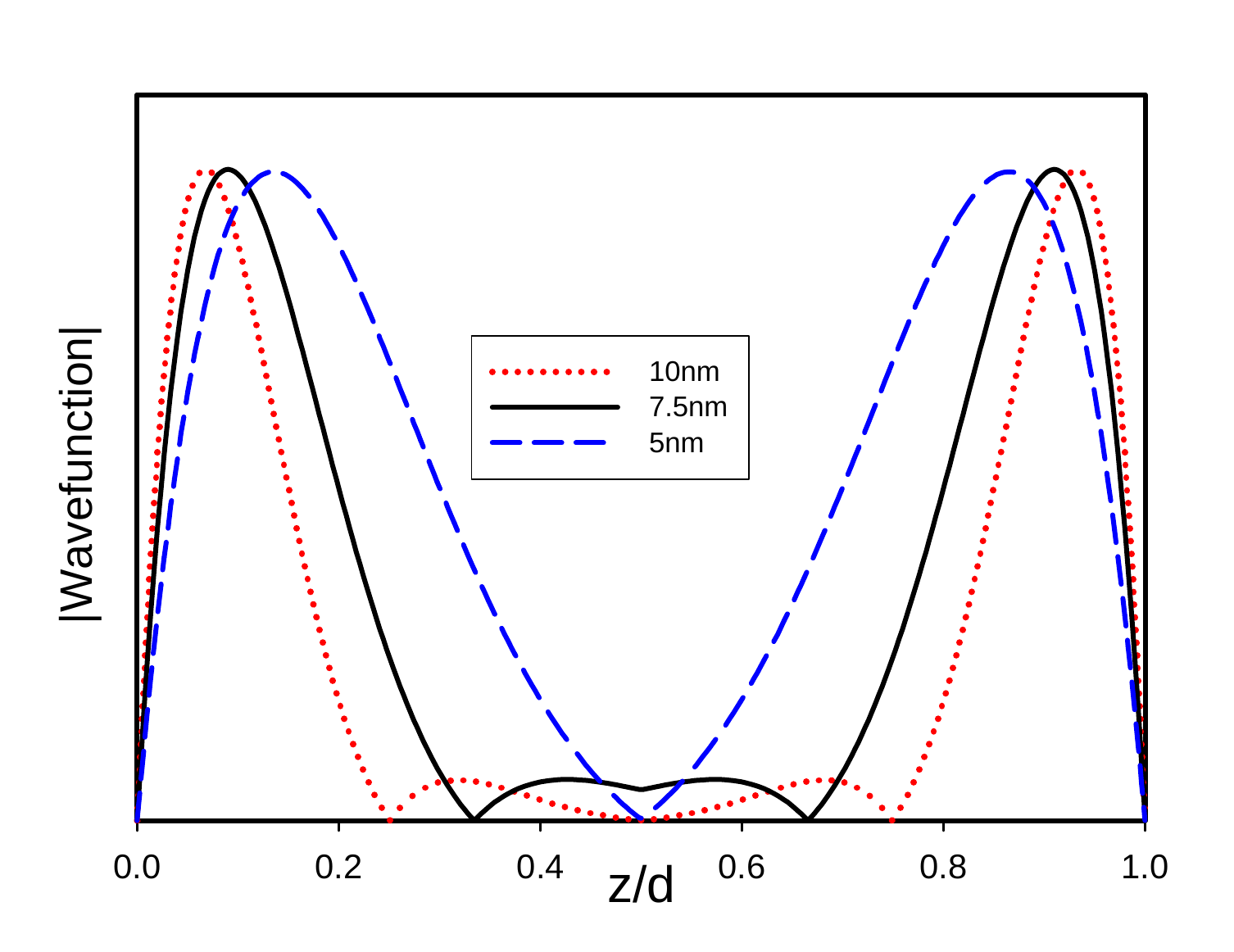}
\caption{(Color online) Absolute value of the wavefunction, as function of the position along the slab, of a surface state with momentum close to zero, for different TI slab thickness.}
\label{wavefunction}
\end{figure}

The energy gap for the Dirac fermions  occurs when the thickness of the TI slab is comparable with the 
decay length of the surface state wave function into the film.  In that case opposite surface wave functions overlap leading to the appearance of the energy gap at the center of the Brillouin zone. In Fig.\ref{wavefunction} we plot the absolute value of the wave function of a surface state with momentum close to zero for different values of the layer thickness. If we turn the parameters $D_1$ and $D_2$ to zero, it is possible to obtain analytically that the four components of the surface wavefunction have the same dependence on the position across  the TI slab,
\begin{equation}
f(z)=e^{\lambda_1 z}\sin {\lambda_2 z} \label{f(z)}
\end{equation}
with
\begin{equation}
\lambda_1+i \lambda _2 = \frac {A_1 + i \sqrt{|A_1 ^2 -4 M_0
B_1|}}{2 B_1} \, \, .
\end{equation}
From the Bi$_2$Se$_3$ band structure parameters we get $\lambda _1
\sim 1.1 nm^{-1}$ and $\lambda _2 \sim 1.26 nm^{-1}$ in rather good agreement with the decays length and zeros of the wavefunction
in Fig.\ref{wavefunction}.  From this values of the decay length, we conclude that in TI slabs thinner than 6nm, there is coupling between top and bottom surfaces and the system can not be described  as a couple of two-dimensional gases separated by a dielectric. It is more appropriated describe the TI slab as a 2D system.

\subsection{Spectrum of a TI slabs in presence of a Exchange field.}

An exchange field directed along  the  $z$-direction affects strongly the electronic  properties of  TI surfaces. The Dirac Hamiltonian describing electrons moving on a  surface perpendicular to the $z$-direction has the form\cite{Liu_2010,Silvestrov_2012} $H$=$\hbar v_F(\sigma _x  k_y -\sigma _y k_x)$.  Here the Pauli matrices $\sigma _x$ and $\sigma_y$ correspond to the electron spin operators and $v_F $ is the Fermi velocity. An exchange field $\Delta _z$ pointing in the $z$-direction opens a gap in the surface bands and the system becomes a topological insulator showing a Hall conductivity $\sigma _{xy}$=$1/2e^2/h$.  The presence of different oriented surfaces in real crystals,  makes that measured Hall conductances become  always integrally quantized\cite{Lee_2009,Vafek_2011,Brey_2014}.

\begin{figure}
\includegraphics[clip,width=8.5cm]{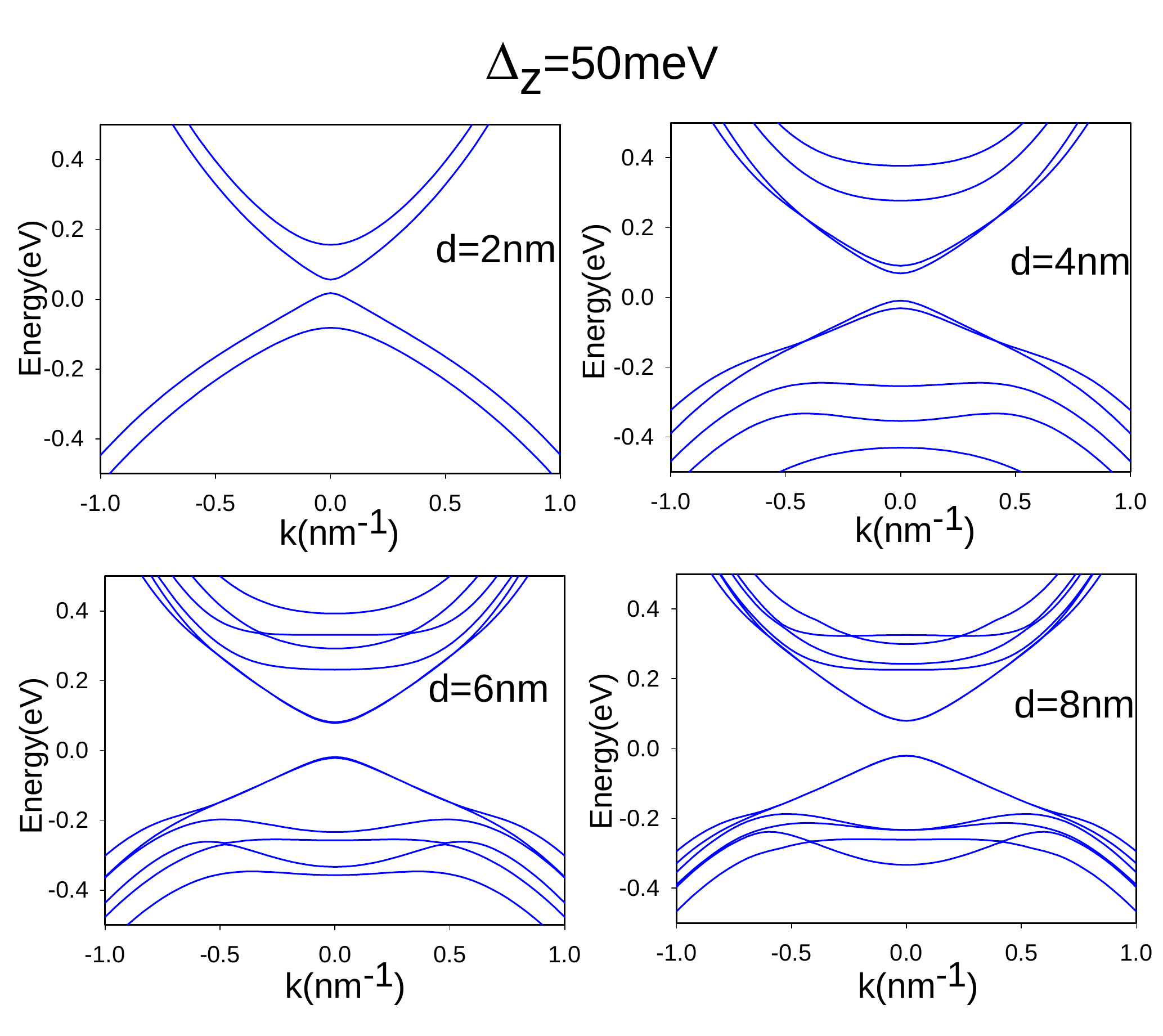}
\caption{(Color online) Band structure of TI slabs of different thickness $d$, in presence of an exchange field, $\Delta_z$=50meV, applied in the $z$-direction. The slabs are perpendicular    to the $\hat z$ direction. The electronic sates are  obtained by diagonalizing Hamiltonian  Eq. \ref{H3D} with the appropriated boundary conditions.  
The system has circular symmetry and we plot the bands as function of the absolute value of the in-plane wavevector ${\bf k}$.}\label{bandasB}
\end{figure}

We study the effect of a exchange field on the properties of a TI thin film, by adding to the bulk Hamiltonian Eq.\ref{H3D} a exchange term of the form
$\Delta _Z I \otimes \sigma _z$. The slab geometry is introduced by  forcing the wavefunctions to satisfy the boundary conditions,  Eq.\ref{2Dwf}.  
In Fig.\ref{bandasB} we plot  the band structure of thin TI slabs for different values of the  thickness $d$ and a typical  exchange field $\Delta _z$=50meV.  

For thick slabs where  opposite surfaces are decoupled, the exchange field polarizes, near ${\bf k}$=0,  the spin of  the conduction(valence)  band  in the $+(-)$ z-direction.
Then a gap of magnitude 
2$\Delta_z$  appears  and the bands get a quadratic dispersion.  Away from the center of the Brillouin zone, the bands recover the linear dispersion and 
the  spin texture dictated by the Dirac equation is restored.

At $\Delta_z$=0,  ultra thin TI slabs have 
two degenerated hyperbolic Dirac bands with  opposite spin texture. The exchange field breaks the time reversal symmetry, and near ${\bf k}$=0 the bands split in two states with spin up and energies 
$\pm E_g/ 2  +\Delta _z$  and two spin down states with energies  $\pm E_g/2  - \Delta _z$.

\section{Optical conductivity}

The optical conductivity $\sigma _{\alpha,\beta}$ relates, in the linear response, the electrical current in a direction $\alpha$ to an external transverse electric
field applied in the $\beta$-direction.

The optical conductivity consists of two pieces, the diamagnetic term and the paramagnetic term\cite{Giuliani:05}. The paramagnetic term can be obtained from the
current-current correlation function using the Kubo formalism. 
However the diamagnetic term is not always well described in continuos effective models. In particular, in the Dirac Hamiltonian the diamagnetic term vanishes   because $\partial ^2 H/\partial k^2$=0 \cite{Stauber10b,Stauber_2013}. 

Both the diamagnetic and paramagnetic contributions contains a delta singularity at frequency $\omega$=0. The total weight of the delta is the Drude weight or charge stiffness and indicates the ability of the carriers to move freely  when an electric field is applied.  In this work we are considering  undoped systems with no carriers at the Fermi energy and therefore the delta function should have null weight. 
To achieve this, we use the following expression\cite{Allen_Book} for the optical conductivity that cancels the delta singularity  at $\omega$=0, and guarantees a zero charge stiffness in the systems,
\begin{widetext}
\begin{equation}
\sigma _{\alpha,\beta} (\omega)= i\frac {e ^2 \hbar }{V} \int \frac { d ^2 {\bf k}}{(2 \pi)^2} \sum _{m,n} \frac { n_F(\varepsilon _{n,{\bf k}} ) -n_F(\varepsilon _{m,{\bf k} } )} {\varepsilon _{m,{\bf k}}-\varepsilon _{n,{\bf k} }   }
\frac {   <n,{\bf k}| j _{\alpha}|m,{\bf k}><m,{\bf k}| j _{\beta}|n,{\bf k}> }
{\hbar (\omega + i \eta) -(\varepsilon _{m,{\bf k}}-\varepsilon _{n,{\bf k} }) }
\label{sigma}
\end{equation}
\end{widetext}
In this expression $|m,{\bf k}>$ and $\varepsilon _{m,{\bf k}}$ are the eigenfunction and eigenvalues  respectively of the system with  sub-band $m$ and wavevector $ {\bf k}$, $n_F$ is the fermi occupation,
and $\hbar \eta$ represents the quasiparticle lifetime broadening.  The current operators are obtained from the Hamiltonian through $j_{\alpha}$=-$\partial H/\partial k _{\alpha}$.

Note that we do not need to include an energy cutoff in the calculation of the optical conductivity. The number of  surface states is naturally limited by the  bulk bands  of the
TI.

\begin{figure}
\includegraphics[clip,width=8.5cm]{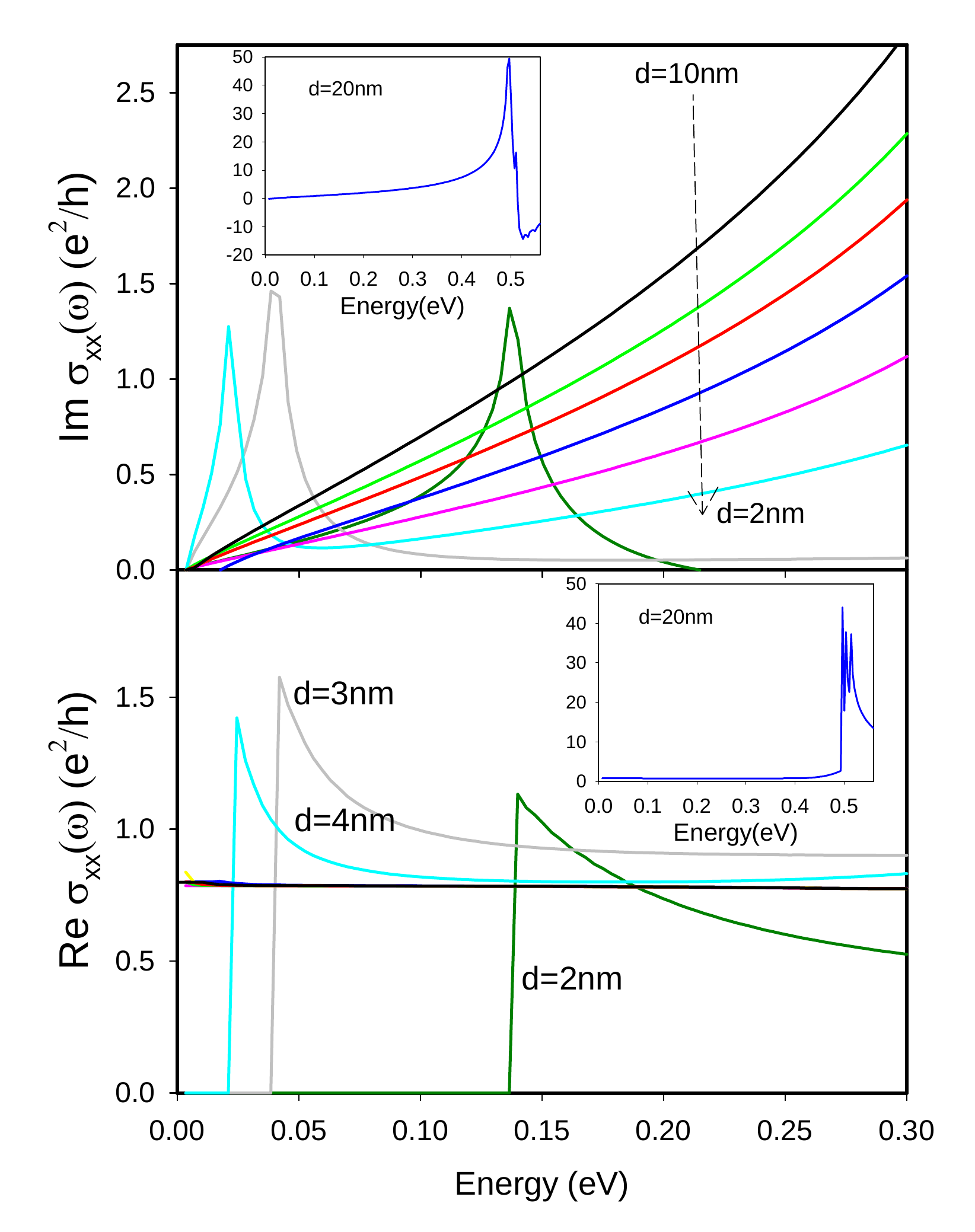}
\caption{(Color online) Imaginary (top)  and real (bottom) part of the optical conductivity as a function of frequency,  for a  TI slab of different thicknesses.  In the insets we show the optical conductivity for a thick slab, $d$=20$nm$ with  states in opposite surfaces practically decoupled}
\label{Sigmaxxb0}
\end{figure}

\subsection{Zero exchange field.}

For zero exchange field the system has time-reversal symmetry and the non-diagonal components of the conductivity tensor are zero. Also because of the  circular symmetry  of the Hamiltonian  the system is isotropic and $\sigma _{xx}$=$\sigma _{yy}$. 
In Fig.\ref{Sigmaxxb0} we show the optical conductivity as a function of frequency for TI slabs of different thickness.  

We first discuss the optical conductivity of uncoupled surfaces. In the insets of Fig.\ref{Sigmaxxb0} we plot the conductivity for a layer thickness $d$=20$nm$.  For this thickness  electronic  states of both surfaces are decoupled  and the dispersion near the center of the Brillouin zone is linear.  Then at small $\omega$,  the imaginary part of the optical absorption is zero and the real part gets the value $\sigma _0$=$\frac {\pi} 4 \frac {e^2}h$. The value of $\sigma _0$ is half the value of the optical conductivity of graphene monolayer\cite{Gusynin_2007,Abedinpour_2011}, because although there is a  Dirac cone at each  surface, they are not spin degenerated.  At frequencies near 0.35eV the dispersion relation deviates from the linear behavior and the real part of the conductivity increases continuously with $\omega$  and presents a strong peak at the TI energy gap.  The imaginary part increases continuously  from zero 
and presents a peak valley structure  at the bulk energy gap. 

At smaller layer thickness the coupling between opposite surface states opens a gap that suppress  the optical absorption at frequencies  smaller than the gap. 
At  $\hbar \omega$=$E_g$,  the hyperbolic dispersion  and the modification of the spin texture near $\Gamma$ increases the number of ${\bf k}$ states that participate in the transition and this produces  a peak superposed to a step  in the real part of  $\sigma _{xx}$. As the gap reduces the step in Re$\sigma_{xx}$ tends to $\sigma _0$ and the peak disappears.
The gap, also induces a peak in the imaginary part of $\sigma _{xx}$  which indicates the existence of an interband  charge collective excitation.

\begin{figure}
\includegraphics[clip,width=8.5cm]{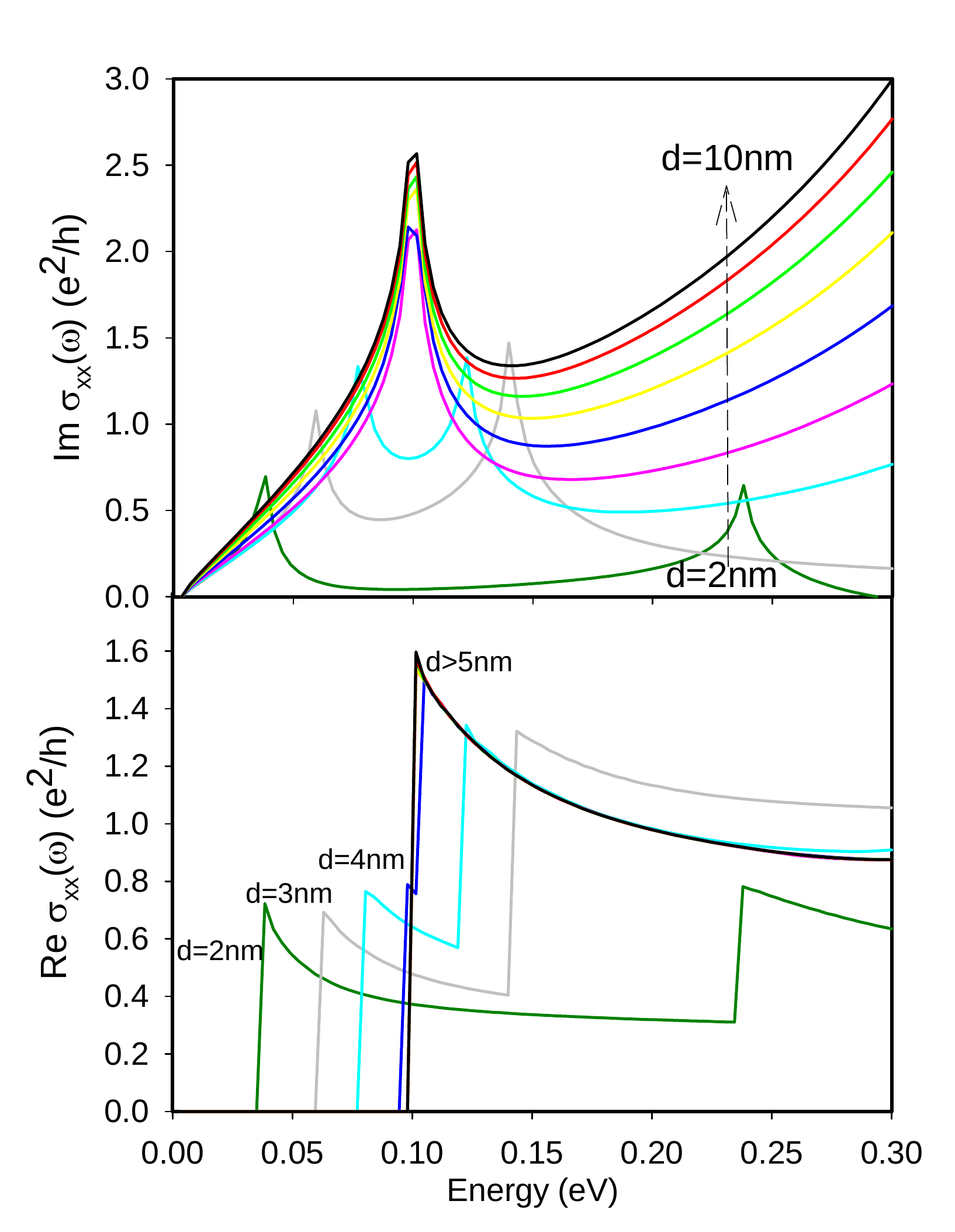}
\caption{(Color online) Imaginary (top)  and real (bottom) part of the longitudinal optical conductivity as a function of frequency,  for a  TI slab of different thicknesses, in presence of an exchange field $\Delta _z$=50meV.}
\label{Sigmaxxb50}
\end{figure}

\begin{figure}
\includegraphics[clip,width=8.5cm]{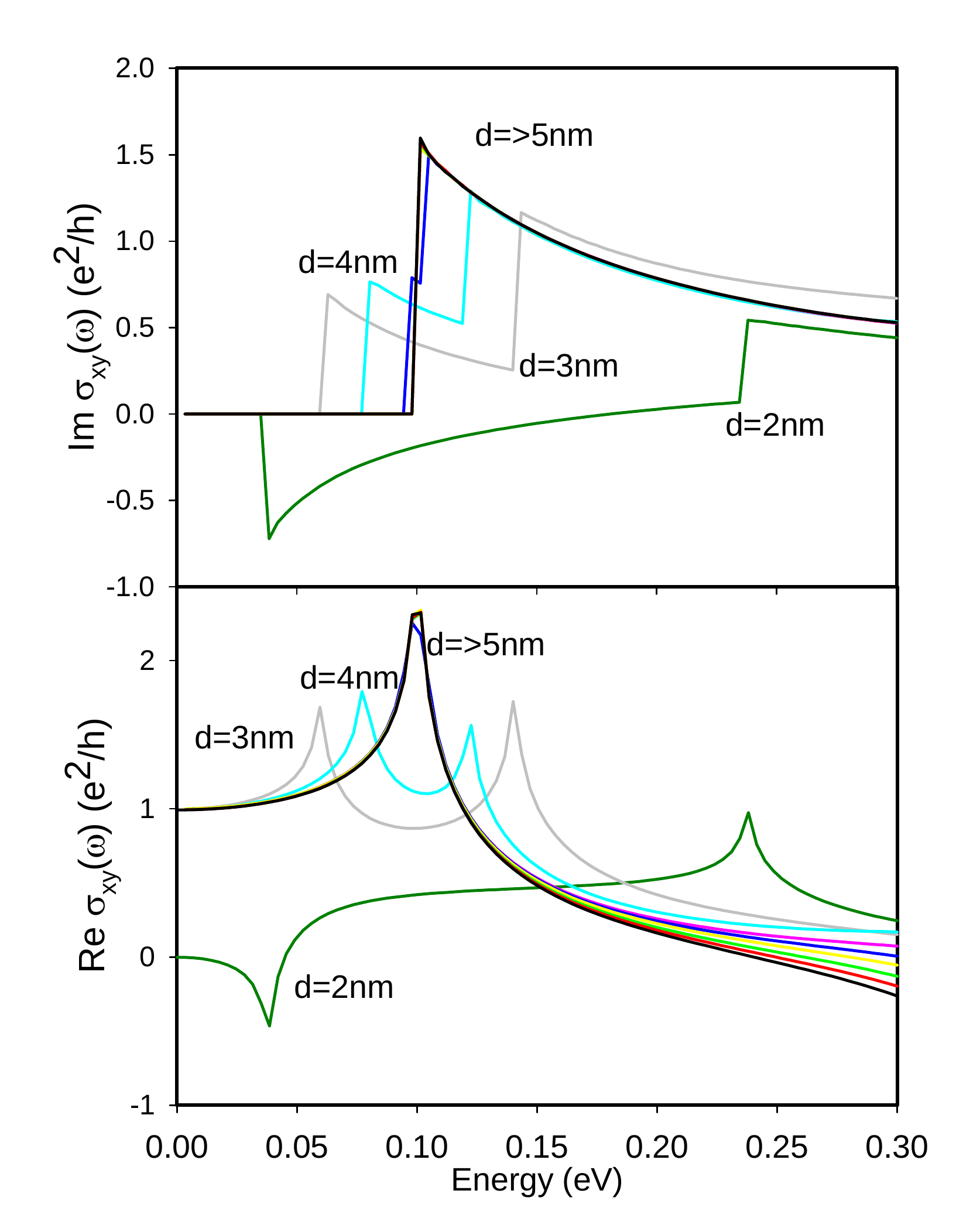}
\caption{(Color online) Imaginary (top)  and real (bottom) part of the Hall optical conductivity as a function of frequency,  for a  TI slab of different  thicknesses, in presence of an exchange field $\Delta _z$=50meV.}
\label{Sigmaxyb50}
\end{figure}

\subsection{Finite exchange field.}\begin{figure}

\includegraphics[clip,width=7.cm]{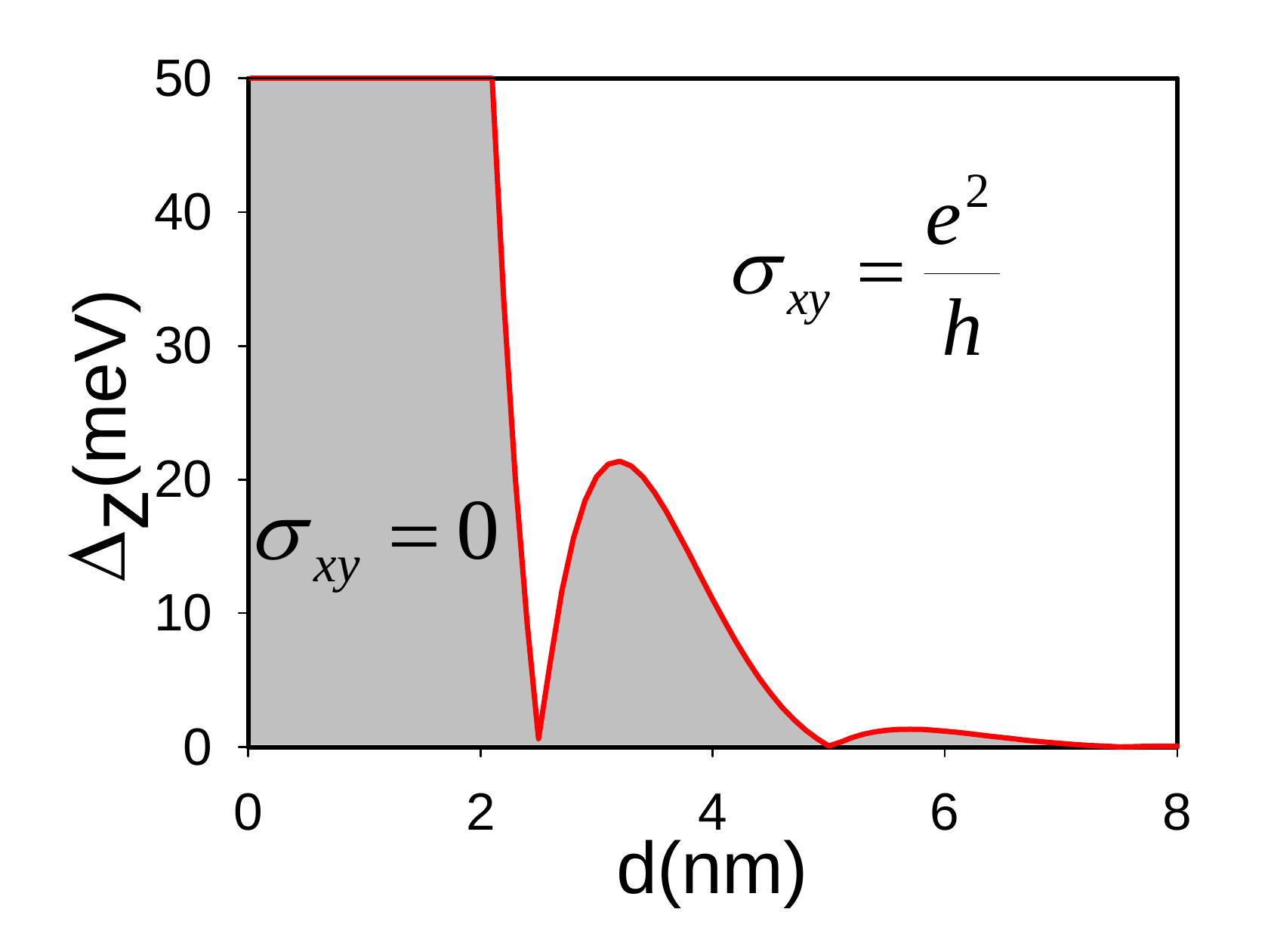}
\caption{(Color online) Phase diagram, as a function of $d$ and $\Delta_z$, for the anomalous quantum Hall effect of a TI slab perpendicular to the $z$-direction.}
\label{PhaseDiagram}
\end{figure}
The exchange field breaks time reversal symmetry and the non-diagonal part of the optical conductivity gets a finite value.
For thin gapped TI slabs, 
the exchange field polarizes the spin in the extremes of the bands,  and the dissipative parts of the response functions,
Re$ \sigma _{xx}$ and Im$\sigma_{xy}$, show two  absorption edges  at energies 2$\Delta_z \pm E_g$. The splitting of the bands  also reflects in the appearance of two peaks in Im$ \sigma _{xx}$ and Re$\sigma_{xy}$. At large separation between the surfaces only an absorption edge and a peak occur at energy 2$\Delta_z$.

A TI slab in presence of a exchange field has always an  insulator character. The band insulator  or topological insulator  nature  of this phase may be known by 
computing   the Chern number, which in this system\cite{hasan_2010,Prada_2011} coincides with the value of the dc  Hall conductivity.
In Fig.\ref{PhaseDiagram} we plot the Hall conductivity as function of $d$ and $\Delta _z$.
For values of the exchange field smaller than twice  the tunneling gap, the slab is a normal insulator, however for larger values
of $\Delta _Z$, the  
TI slab  acquires an integer  anomalous Hall conductivity, $\sigma _{xy}$=$e^2 /h$. 
Then as a function of the  exchange field or the slab thickness, TI slabs undergo  quantum phase transitions from band to topological insulator.
The system only presents an anomalous quantum Hall effect when the surface bands are inverted by the exchange field, independently of the sign of the tunneling gap between the surfaces.
For very large values of $d$, the integer quantized Hall conductivity may be understood as  the sum of  two half quantized Hall conductivities, one for each uncoupled surface.

\section{Kerr and Faraday angles.}

\begin{figure}
\includegraphics[clip,width=8.5cm]{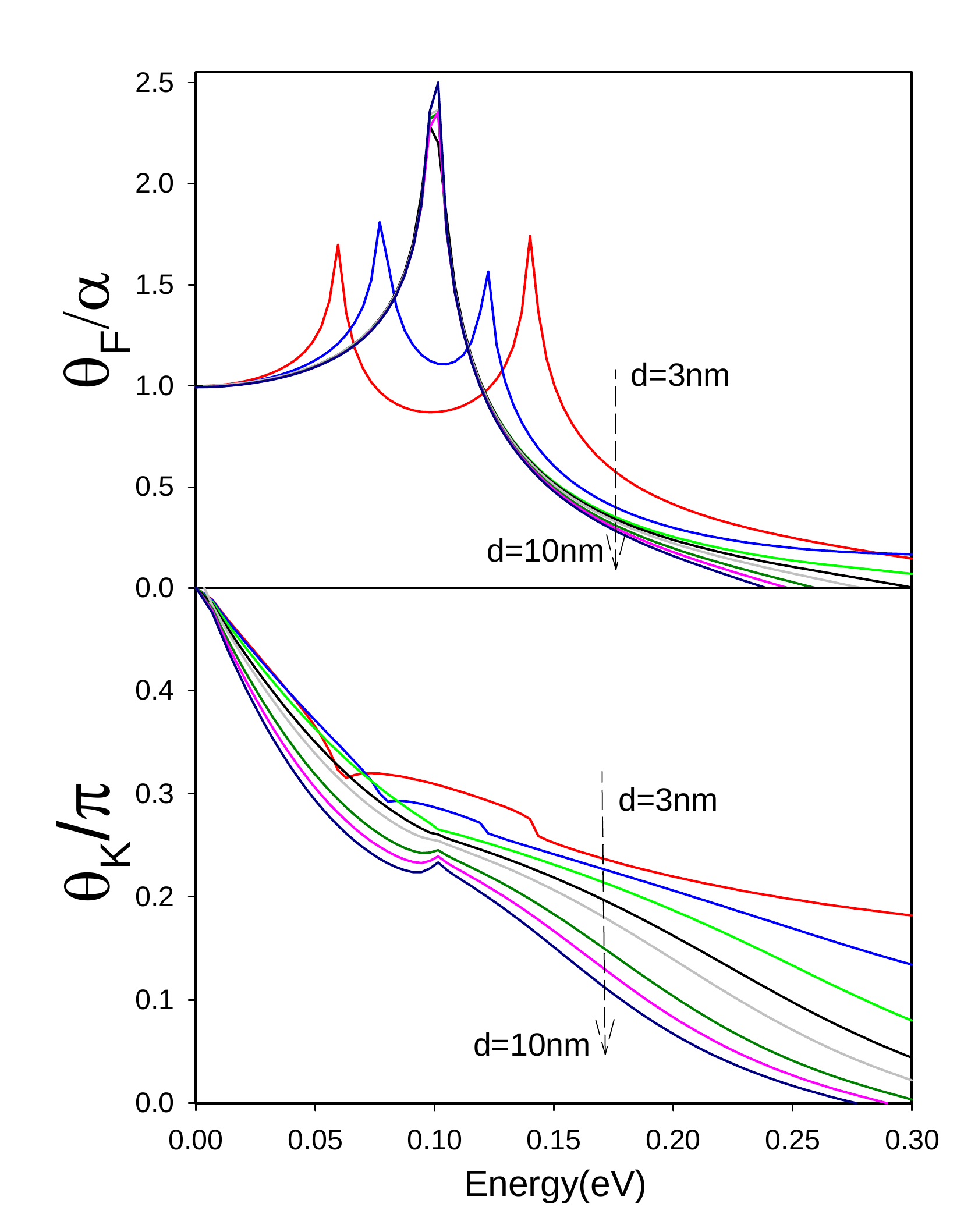}
\caption{(Color online) Faraday and Kerr angles for TI slabs of different thicknesses in presence of an exchange field $\Delta _z$=50meV.}
\label{Angulos}
\end{figure}

Finite values of the Hall conductivity in magnetically gapped TI surfaces imply interesting properties in their optical properties. When $\sigma_{xy}\ne$0, left and right handed circularly polarized light, propagating perpendicular to the TI slab, have different refraction and transmission indices. As a consequence,  linear polarized light rotates its polarization direction when transmitted or reflected by a TI slab in the anomalous quantum Hall regime.
The transmitted and reflected rotation angles of linear polarized light are called Kerr ($\theta _K$) and Faraday ($\theta _F$) respectively. 

Because of the overlap between  electronic states  localized on opposite surfaces,  ultra thin TI slabs should be considered  as  a single two dimensional electron system. Furthermore, the light wavelength is much larger than the slab thickness and the light electromagnetic fields   are practically constants across the slab.  In this situation, the electromagnetic properties of the thin TI slab are described just   by the two dimensional conductivity tensor. 
Transmission and reflection of light   are obtained  by  considering two dielectric materials separated by an  interface characterized by a conductivity tensor. 

Maxwell equations dictates the boundary conditions for the electromagnetic fields. For a free standing  slab and normal incidence, the
reflection, $ {\bar r }$=$\left ( \begin{array}{cc} r_{xx} & r_{xy} \\ -r_{xy} & r _{xx} \end{array} \right )$ and transmission,
$ {\bar t }$=$\left ( \begin{array}{cc}t_{xx} & t_{xy} \\ -t_{xy} & t _{xx} \end{array} \right )$ tensors for the electric field have the form\cite{Tse_2011},
\begin{eqnarray}
r_{xx}& = &\frac {1-(1+\frac{4 \pi} c \sigma _{xx}) ^2 -(\frac {4 \pi} c \sigma _{xy}) ^2}
{(2+\frac {4 \pi } c \sigma _{xx} ) ^2 +(\frac {4 \pi} c \sigma _{xy})^2} \nonumber \\
r_{xy} & =  &\frac {-\frac {8 \pi} c \sigma _{xy}}
{(2+\frac {4 \pi } c \sigma _{xx} ) ^2 +(\frac {4 \pi} c \sigma _{xy})^2} \nonumber \\
t_{xx} & = & \frac {4+\frac{8 \pi} c \sigma _{xx} }
{(2+\frac {4 \pi } c \sigma _{xx} ) ^2 +(\frac {4 \pi} c \sigma _{xy})^2} \nonumber \\
t_{xy} & =  & \frac {-\frac {8 \pi} c \sigma _{xy}}
{(2+\frac {4 \pi } c \sigma _{xx} ) ^2 +(\frac {4 \pi} c \sigma _{xy})^2} \, .
\label{retr}
\end{eqnarray}
and the Kerr and Faraday angles are given by,
\begin{eqnarray}
\theta _F & = &{\rm arg} (t_{xx}+i t_{xy}) \nonumber \\
\theta _K &= & {\rm arg} (r_{xx}+i r_{xy} )
\end{eqnarray}

In Fig.\ref{Angulos} we plot the Kerr and Faraday angles for TI slabs with thickness ranging from $d$=2nm to $d$=10nm, in presence of an exchange field $\Delta_z$=50meV. 
For $\Delta _z  \! \ne$0, TI slabs exhibit  anomalous quantum Hall effect characterized by $\sigma _{xx} (\omega \! =\! 0) $=0, Re$\sigma _{xy} (\omega \! =\! 0)$=0 and Im$\sigma_{xy}(\omega \! =\! 0)$=$e^2/h$, then the Kerr angle, Eq.\ref{retr}, gets a large (giant) value $\theta_K$=-$\pi/2$ and the Faraday gets a quantized value $\theta_F$=$\alpha$, being $\alpha$=$e^2/\hbar c$=1/137 the vacuum fine structure constant\cite{Tse_2010,Tse_2011}. Absolute values of both angles decay with frequency  and present peaks at the optical absorption edges $\omega$=$2\Delta_z \pm E_g$.  

The results indicate that  in a wide range of frequencies, linear polarized light gets a large rotation in its polarization when reflected or transmitted by 
an ultra thin TI slab. The small value of the slab thickness makes this result rather robust.
For ultra-thin TI slabs the bulk contribution to the electronic conductivity is totally suppressed and the values of $\theta_K$ and $\theta_F$ are not affected by bulk free carriers supply by bulk defects. Also because  the thickness of the slab is much smaller than the light  wavelength,
the optical path of the light inside the  TI slab is zero and there is not suppression of $\theta _K$ and $\theta _F$ by multiple reflections in side the TI slab. We then conclude that in ultra-thin  TI slabs there is a wide range of frequencies where the Kerr and Faraday angles get large values.

\section{Conclusions.}
We have studied the band structure and optical properties of ultra-thin topological insulator slabs. The electronic properties are obtained starting from a three-dimensional ${\bf k} \cdot {\bf p}$ Hamiltonian, so that our calculations do not depend on energy or momentum cutoff.
For the optical conductivity we use an expression (Eq.\ref{sigma}) that  describes correctly both the diamagnetic and the paramagnetic contributions, and therefore we obtain the real and imaginary part of the conductivity directly from it. 

In thin TI slabs, the coupling between opposite surface states  opens a gap in the electronic spectrum that inhibits 
optical absorption for frequencies smaller than the gap $E_g$. The gap also reflects in a   peak  at $\hbar \omega$=$E_g$ in the imaginary part of  $\sigma _{xx} (\omega)$. This peak
indicates the existence of an interband charge excitation.

An exchange field $\Delta_z$ applied perpendicularly to the TI slab, breaks time-revesal symmetry and splits the gapped bands of ultra-thin TI slabs. The real part of the longitudinal 
optical conductivity  shows  absorption edges  at energies  $2\Delta_z \pm E_g$, whereas the imaginary part presents peaks at the same energies. For values of the exchange field $2\Delta _z >E_g$, the zero frequency Hall conductivity is quantized to the value
$\sigma _{xy}$=$e^2/h$.  This large value of the Hall conductivity produces a strong rotation of the polarization of light when transmited  through or reflected in the TI slab.  At zero frequency the Faraday angle gets the value $\theta _F$=$e^2/\hbar c$=1/137 and presents peaks 
at frequencies $2\Delta_z \pm E_g$. The Kerr rotation gets a value $\theta_K$=-$\pi/2$  at zero frequency 
and gets rather large absolute values at finite frequencies.  These values of the Faraday and Kerr rotation angles are
rather robust because they are not affectd by bulk carriers or by light multiple scattering in the topological insulator ultra-thin slab.

\section{acknowledgments.}
Funding for this work was provided by MEC-Spain via grant FIS2012-33521.


\end{document}